\documentclass[
amsmath,
amssymb,
notitlepage,
aps,
pra,
citeautoscript,
reprint
]{revtex4-1}

\usepackage[greek, ngerman, english]{babel}
\usepackage{graphicx}
\usepackage{siunitx}
\usepackage{hyperref}
\usepackage{bbold}
\usepackage{xcolor}

\newcommand{\upgreek}[1]{\text{{\greektext #1}}}

\newcommand{\upmu}{\upgreek{m}}

\DeclareSIUnit{\Electronmass}{\text{\ensuremath{m_{0}}}}

\begin{document}


\title{Excitonic terahertz absorption in semiconductors with effective-mass anisotropies}
\date{\today}
\author{P.~Springer}
\email{phillip.springer@physik.uni-marburg.de}
\affiliation{Department of Physics and Material Sciences Center, Philipps-Universit\"at Marburg, Renthof 5, 35032 Marburg, Germany}
\author{S.~W.~Koch}
\affiliation{Department of Physics and Material Sciences Center, Philipps-Universit\"at Marburg, Renthof 5, 35032 Marburg, Germany}
\author{M.~Kira}
\affiliation{Department of Physics and Material Sciences Center, Philipps-Universit\"at Marburg, Renthof 5, 35032 Marburg, Germany}


\begin{abstract}
A microscopic approach is developed to compute the excitonic properties and the corresponding terahertz response for semiconductors characterized by anisotropic effective masses.
The approach is illustrated for the example of germanium where it is shown that the anisotropic electron mass in the \textit{L}-valley leads to two distinct terahertz absorption resonances separated by $\SI{0.8}{\milli\electronvolt}$.
\end{abstract}


\maketitle


\section{Introduction}

Terahertz (THz) spectroscopy has been broadly applied, e.g., to investigate transient photoconductivity~\cite{Beard2000}, inter-molecular vibrations~\cite{Nagai2005}, high-harmonic generation~\cite{Schubert2014}, and the transition energies between excitonic eigenstates in quantum many-body systems~\cite{Groeneveld1994,Cerne1996,Kaindl2003}.
For direct-gap semiconductors with isotropic effective-mass configurations, the excitonic THz excitation dynamics and the resulting spectra have been studied extensively both theoretically~\cite{Chao1991,Mathieu1992,Chen2001,Piermarocchi2002} and experimentally~\cite{Pantke1992,Bacher1993,Vinattieri1994,Finkelstein1998}.

In comparison to direct-gap systems, the corresponding investigations in indirect semiconductors such as silicon (Si) and germanium (Ge) are more elaborate because optical excitations are accompanied by strong dephasing due to intervalley scattering~\cite{Lange2009} and the indirect excitons typically involve states characterized by strongly anisotropic masses~\cite{Nathan1963}.
Experimentally, excitonic features have been observed in Ge~\cite{Zwerdling1959} and Si~\cite{Laude1971} and THz studies have been reported recently~\cite{Suzuki2009,Suzuki2012}.

Whereas most of the isotropic exciton properties can be determined analytically~\cite{Mathieu1992,Chao1991}, even the linear eigenvalue problem must be solved numerically for anisotropic conditions.
These subtleties complicate the microscopic analysis of the linear and nonlinear optical experiments, and in particular also of the THz absorption measurements.  

To deal with this problem, we develop in this paper a microscopic approach and an ensuing numerical scheme to efficiently evaluate the excitonic properties in systems with anisotropic effective masses.
To illustrate the scheme, we analyze Ge and show that the THz absorption exhibits distinct resonances related to the $L$-valley electron-mass anisotropy. 

The paper is organized as follows: In Sec.~\ref{sec:theory}, we extend the generalized Wannier equation to systems with mass anisotropy and discuss the system Hamiltonian and basic THz absorption equations.
In Sec.~\ref{sec:Radial_Eigenvalue_Problem}, we present an efficient numerical scheme to obtain the radial solutions of the Wannier equation.
We analyze the modifications of the selection rules and the THz absorption spectra for different polarizations in Sec.~\ref{sec:analysis}.
For the example of Ge, we then show that the mass anisotropy results in two clearly separated exciton resonances in the THz absorption spectrum.


\section{Theory with Mass Anisotropy}
\label{sec:theory}

\begin{figure}[tbp]
\includegraphics{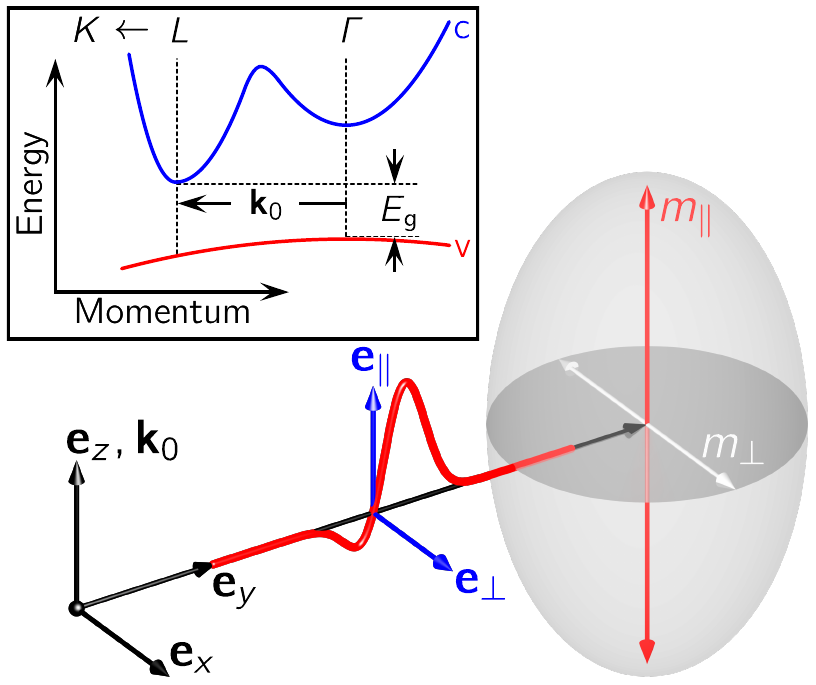}
\caption{
Schematic illustration of the system.
In the $xy$-plane (dark grey), the effective electron mass is denoted by $m_{\perp}$ (white arrow), in $z$-direction it is $m_\parallel$ (red arrow).
The THz field (red) is either polarized parallel or perpendicular (blue arrows) to the $z$-axis which is aligned with ${\bf k}_0$.
The inset schematically depicts the band structure of Ge.
}
\label{fig:schematic}
\end{figure}

To identify the main consequences of mass anisotropy in the THz absorption spectra, we use Ge as a prototype system and a 2-band model to describe the energy dispersion.
Germanium is an indirect semiconductor whose conduction (c) and valence (v) bands are centered around the $L$ and $\varGamma$ points, respectively, separated by the wave vector ${\bf k}_0$, as indicated in the inset of Fig.~\ref{fig:schematic}.
For excitations close to the band gap $E_\mathrm{g}$, it is sufficient to describe the electronic energies via~\cite{Rossler2001},
\begin{align}
\label{eq:eff_mass_appr}
E^{\mathrm{v}}_{{\bf k}} & = -E^{\mathrm{h}}_{{\bf k}} = - \frac{\hbar^2 k^2}{2m_{\mathrm{h}}} \, , \\
E^{\mathrm{c}}_{{\bf k}} & = E^{\mathrm{e}}_{{\bf k}} =
E_\mathrm{g} + \sum_j \frac{\hbar^2 \left[ ({\bf k} - {\bf k}_0 ) \cdot {\bf e}_j \right]^2}{2 m_{\mathrm{e},j}} \, ,
\end{align}
for the holes (h) and the electrons (e), respectively.
The indices $j=\{x,y,z\}$ denote the Cartesian components and ${\bf k}_0$ is aligned with the ${\bf e}_z$ direction, as shown in Fig.~\ref{fig:schematic}.
Although ${\bf k}_0$ in general defines a group of energy minima, we first evaluate the theory for one single ${\bf k}_0$ and generalize the results for multiple ${\bf k}_0$ in Sec.~\ref{sec:multi_valley}.
In Ge, the group of ${\bf k}_0$ points to the eight $L$ centers.
These lie in the center of the hexagonal planes of the truncated octahedron which defines the first Brillouin zone.

In Ge, the valence band is isotropic with mass $m_\mathrm{h} = \SI{0.33}{\Electronmass}$ while the conduction band masses are $m_{\mathrm{e},x} = m_{\mathrm{e},y} \equiv m_\perp = \SI{0.0815}{\Electronmass}$ and $m_{\mathrm{e},z} \equiv m_\parallel = \SI{1.59}{\Electronmass}$ at the $L$ valley~\cite{Madelung1991}.
Figure~\ref{fig:schematic} illustrates the directions of the anisotropic effective mass tensor as an ellipsoid with the $xy$-plane mass given by $m_\perp$ and with $m_\parallel$ in the $z$-direction.

\subsection{System Hamiltonian}\label{sec:System_Hamiltonian}

The many-body Hamiltonian is $\hat{H}=\hat{H}_0+\hat{V}+\hat{H}_{\mathrm{THz}}$ where the non-interacting electrons are described by~\cite{Kira2006}
\begin{align}
\hat{H}_0 = \sum_{{\bf k}} \left[
E^\mathrm{c}_{{\bf k}} \hat{a}^\dagger_{\mathrm{c},{\bf k}}\hat{a}_{\mathrm{c},{\bf k}}
+
E^\mathrm{v}_{{\bf k}} \hat{a}^\dagger_{\mathrm{v},{\bf k}}\hat{a}_{\mathrm{v},{\bf k}}
\right] \, ,
\end{align}
with Fermionic creation (annihilation) operators $\hat{a}^\dagger_{\lambda,{\bf k}}$ ($\hat{a}_{\lambda,{\bf k}}$) for conduction ($\lambda = \mathrm{c}$) and valence band ($\lambda = \mathrm{v}$), respectively. 
The Coulomb-interaction is given by~\cite{Haug2009a}
\begin{align}
\hat{V} = \frac{1}{2}
\sum_{\lambda,\nu}\sum_{{\bf k},{\bf k}'\!,{\bf q}}
V_q
\hat{a}^\dagger_{\lambda,{\bf k}}
\hat{a}^\dagger_{\nu,{\bf k}'}
\hat{a}_{\nu,{\bf k}'+{\bf q}}
\hat{a}_{\lambda,{\bf k}-{\bf q}}
\, ,
\end{align}
containing the usual Coulomb matrix element $V_q$.
For weak THz fields, the light-matter coupling follows from~\cite{Kira2003}
\begin{align}
\label{eq:THz_inter}
\hat{H}_\mathrm{THz}
=
- \sum_{\lambda,{\bf k}}
{\bf j}^\lambda_{{\bf k}} \cdot {\bf A}_\mathrm{THz}
\hat{a}^{\dagger}_{\lambda,{\bf k}}
\hat{a}_{\lambda,{\bf k}} \, ,
\end{align}
with the current-matrix elements in the effective-mass approximation,
\begin{align}
\label{eq:current_matrix_element}
{\bf j}^\mathrm{h}_{{\bf k}}
=
- \frac{|e|\hbar{\bf k}}{m_\mathrm{h}}
\, ,
\quad
{\bf j}^\mathrm{e}_{{\bf k}}
=
-|e| \hbar \sum_j
\frac{
({\bf k} - {\bf k}_0) \cdot {\bf e}_j
}{
m_{\mathrm{e},j}
} {\bf e}_j \, ,
\end{align}
and a THz field ${\bf A}_\mathrm{THz}(t) \equiv A(t) {\bf e}_A$.
Due to the mass anisotropy in ${\bf j}_{{\bf k}}^{\mathrm{e}}$, the THz interaction is sensitive to the polarization ${\bf e}_A$ of the applied field.


\subsection{Anisotropic Excitons}\label{sec:Anisotropic_Excitons}

To compute the THz probe absorption spectrum, we have to specify the initial many-body state of the semiconductor.
Here, we consider a situation where the $L$-point electrons and the $\varGamma$-point holes have formed bound electron--hole pairs, i.e.~excitons~\cite{Knox1963}.
The corresponding exciton states $\phi_\lambda$ and their binding energies $E_\lambda$ have to be computed from the generalized Wannier equation~\cite{Haug1984,Kira2006}
\begin{align}
\label{eq:Wannier}
E_\lambda \phi_\lambda^{\mathrm{R}}({\bf k})
& =
\tilde{E}_{\bf k} \phi_\lambda^{\mathrm{R}}({\bf k})
 - \left(1 - f^{\mathrm{e}}_{{\bf k}} - f^{\mathrm{h}}_{{\bf k}}\right)
\sum\limits_{{\bf k}'}
V_{|{\bf k} - {\bf k}'|} \phi_\lambda^{\mathrm{R}}({\bf k}') \, ,
\end{align}
where $f^{\mathrm{e}(\mathrm{h})}_{{\bf k}}$ is the electron (hole) distribution.
Further Coulomb correlation effects, such as excitation induced dephasing~\cite{Schmitt-Rink1989,Wang1993}, could be included via complex scattering matrices~\cite{Hoyer2006}, but are omitted here for simplicity.

Non-vanishing carrier distributions renormalize the electron--hole pair energy
\begin{align}
\label{eq:ene_disp_aniso}
\tilde{E}_{{\bf k}} =
\sum_j \frac{
\hbar^2 \left[ {\bf k} \cdot {\bf e}_j \right]^2
}{2 \mu_j}
- \sum\limits_{{\bf k}'}
V_{|{\bf k} - {\bf k}'|}
\left(
f^{\mathrm{e}}_{{\bf k}'} + f^{\mathrm{h}}_{{\bf k}'}
\right) \, ,
\end{align}
after we have introduced a reduced mass $\mu_{j}^{-1} = m_{e,j}^{-1} + m_h^{-1}$.
Since two of the three $\mu_j$ are identical in Ge, the energy dispersion~(\ref{eq:ene_disp_aniso}) simplifies to
\begin{align}
\label{eq:ene_disp_aniso_special}
\tilde{E}_{{\bf k}} =
\frac{\hbar^2 k_{\perp}^2}{2\mu_{\perp}}
+ \frac{\hbar^2 k_{\parallel}^2}{2\mu_{\parallel}}
- \sum\limits_{{\bf k}'} V_{|{\bf k} - {\bf k}'|}
\left(
f^{\mathrm{e}}_{{\bf k}'} + f^{\mathrm{h}}_{{\bf k}'}
\right) \, ,
\end{align}
with $\mu_{\perp(\parallel)}^{-1}=m_{\perp(\parallel)}^{-1} + m_\mathrm{h}^{-1}$ and the momentum ${\bf k}=({\bf k}_\perp,k_\parallel)$, both being decomposed into directions perpendicular ($\perp$) and parallel ($\parallel$) to ${\bf k}_0$, as shown in Fig.~\ref{fig:schematic}.

In general, $\tilde{E}_{\bf k}$ is anisotropic for $\mu_\perp \neq \mu_\parallel$.
For non-vanishing carrier distributions, the Wannier equation defines a non-Hermitian eigenvalue problem with left- and right-handed solutions $\phi_\lambda^{\mathrm{L}}$ and $\phi_\lambda^{\mathrm{R}}$, respectively.
As shown in Ref.~\cite{Kira2012}, these solutions are connected via $
\phi^{\mathrm{L}}_\lambda({\bf k}) = \phi^{\mathrm{R}}_\lambda({\bf k})/(1 - f^{\mathrm{e}}_{{\bf k}} - f^{\mathrm{h}}_{{\bf k}})
$.
Due to the mass anisotropy and the $f^\lambda_{\bf k}$ dependence, Eq.~(\ref{eq:Wannier}) cannot be solved analytically.
In Sec.~\ref{sec:Radial_Eigenvalue_Problem}, we therefore present a method to numerically determine the anisotropic exciton wave functions $\phi^\mathrm{R}_\lambda$.

Once the exciton wave functions are known, we can directly evaluate the THz absorption via the susceptibility
\begin{align}
\label{eq:THz_Elliott}
\chi(\omega)
&=
\sum_{\lambda,\nu}
\frac{
S^\nu_\lambda (\omega) n_\lambda^\nu - \left[
S^\nu_\lambda(-\omega) n_\lambda^\nu
\right]^\star
}{\epsilon_0 \omega^2 (\hbar\omega + \mathrm{i} \gamma_J)
}
\, ,
\end{align}
derived in Ref.~\cite{Kira2006}.
The susceptibility defines the linear absorption $\alpha(\omega) = \omega/c_\mathrm{r}\mathrm{Im}[\chi(\omega)]$ yielding the THz Elliott formula, where $c_\mathrm{r}$ is the speed of light within the medium.
Equation~(\ref{eq:THz_Elliott}) also contains a decay constant $\gamma_J$ for the THz current, as well as a THz response function
\begin{align}
\label{eq:response_function}
S_\lambda^\nu (\omega) =
\sum_\beta
\frac{
(E_\beta-E_\lambda)J_\lambda^\beta J_\beta^\nu
}{
E_\beta-E_\lambda-\hbar\omega-\mathrm{i}\gamma
} \, ,
\end{align}
where $\gamma$ is the dephasing constant, and $n_\lambda^\nu$ assign exciton correlations.
The transition-matrix element
\begin{align}
\label{eq:THZ_trans_MEL}
J_\lambda^\nu =
\sum_{{\bf k}}
\left[
\phi^{\mathrm{L}}_\lambda({\bf k})
\right]^\star
j({\bf k}) \phi^{\mathrm{R}}_\nu({\bf k})
\end{align}
contains the reduced current
$
j({\bf k}) = \left({\bf j}^\mathrm{e}_{{\bf k}} + {\bf j}^\mathrm{h}_{{\bf k}} \right) \cdot {\bf e}_A
$.
Since both the Wannier equation and $j({\bf k})$ are generally anisotropic in systems with anisotropic effective masses, the THz absorption will depend on the direction of detection, which allows us to directly monitor the mass anisotropy via $\alpha(\omega)$.


\begin{figure}[tbp]
\includegraphics[]{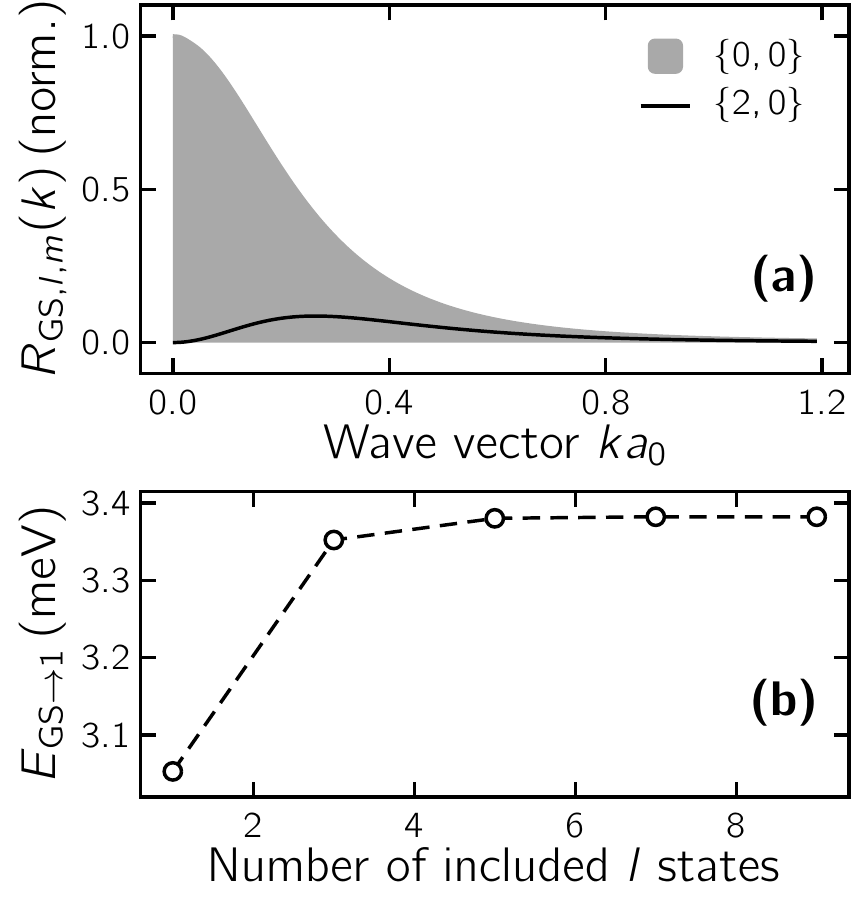}
\caption{
Frame (a) shows the different radial parts $R_{\mathrm{GS},l,m}$ of the ground state whose dominant character is $\{l=0,m=0\}$ (shaded area).
The only other non-vanishing component has $\{2,0\}$ symmetry (solid line).
The Bohr radius is $a_0=\SI{30.98}{\angstrom}$.
Frame (b) shows the computed transition energy $E_{\mathrm{GS} \to 1}$ (circles) between the two energetically lowest states as a function of maximally included $l$ states.
The dashed line acts as a guide to the eye.
}
\label{fig:numerics}
\end{figure}

\section{Anisotropic Eigenvalue Problem}\label{sec:Radial_Eigenvalue_Problem}

In the case of isotropic effective masses, the radial and angular dependency in the Wannier equation can be separated such that Eq.~(\ref{eq:Wannier}) can be solved as an effective one dimensional problem.~\cite{Schafer2006,Hoyer2006,Steiner2008a}.
However, since the rotation symmetry is broken by the mass anisotropy, the radial and angular degrees of freedom are coupled leading to a significantly more complicated configuration. 

In the following, we develop an efficient scheme to deal with the general solution of the Wannier equation in anisotropic media.
For this purpose, we allow for all $\mu_j$ in Eq.~(\ref{eq:ene_disp_aniso}) to be different.
In the case of Ge, we then simply take two of the masses as equal. 

First of all, we expand $\phi^\mathrm{R}_\lambda$ into spherical harmonics
\begin{align}
\label{eq:WF_ansatz}
\phi_{\lambda}^\mathrm{R} ({\bf k}) = \sum_{l=0}^\infty\sum_{m=-l}^l R_{\lambda,l,m}(k) Y_l^m(\theta, \varphi) \, .
\end{align}
Inserting Eq.~(\ref{eq:WF_ansatz}) into Eq.~(\ref{eq:Wannier}) and projecting spherical harmonics yields an eigenvalue problem for the radial part alone
\begin{align}
\label{eq:radial_Wannier}
E_{\lambda} R_{\lambda,l,m}(k) & =
\frac{\hbar^2k^2}{2\mu_z} \epsilon_{l,m}^{(1)}R_{\lambda,l,m}(k) \nonumber \\
&- \int\limits_0^\infty \mathrm{d}k' \left(k'\right)^2 \mathcal{V}^0_{k,k'} \left(f^{\mathrm{e}}_{k'} + f^{\mathrm{h}}_{k'}\right)R_{\lambda,l,m}(k) \nonumber \\
&- \left(1 - f^{\mathrm{e}}_{k} - f^{\mathrm{h}}_{k}\right) \int\limits_0^\infty \mathrm{d}k' \left(k'\right)^2 \mathcal{V}^l_{k,k'} R_{\lambda,l,m}(k')
\nonumber \\
& + \frac{\hbar^2k^2}{2\mu_z} \sum\limits_{\xi=\pm1}
\Bigg[
\epsilon_{l+\xi,m}^{(2)} R_{\lambda,l+2\xi,m}(k) \nonumber \\
&+ \epsilon_{l,m+\xi}^{(3)} R_{\lambda,l,m+2\xi}(k) \nonumber \\
&+ \sum\limits_{\sigma=\pm1} \epsilon_{l+\sigma,\xi \sigma m + (\sigma-1)}^{(4)} R_{\lambda,l+2\sigma,m+2\xi}(k)
\Bigg]
\, ,
\end{align}
when $f^{\mathrm{e}(\mathrm{h})}_{\bf k} = f^{\mathrm{e}(\mathrm{h})}_k$ is radially symmetric.
The mass anisotropy introduces coupling strengths $\epsilon^{(j)}_{l,m}$ and Coulomb-matrix elements $\mathcal{V}^l_{k,k'}$ given in App.~\ref{sec:Coupling_Energies} and \ref{sec:Coulomb_Matrix_Element}, respectively.
If the carrier distributions are not radially symmetric, they can be expanded similarly to Eq.~(\ref{eq:WF_ansatz}).

Equation~(\ref{eq:radial_Wannier}) couples only identical $l$ and $m$ elements and those shifted by $\pm1$.
At the same time, the expansion~(\ref{eq:WF_ansatz}) typically provides a fast convergence in terms of included $l$ states.
Therefore, this eigenvalue problem can be solved with only a few $l$ and $m$ combinations.
For isotropic electron masses, Eq.~(\ref{eq:radial_Wannier}) reduces to the usual form of the radial Wannier equation.

For Ge-type systems, Eq.~(\ref{eq:radial_Wannier}) simplifies significantly because $\epsilon^{(3)}_{l,m} = \epsilon^{(4)}_{l,m} = 0$ for two identical masses, as shown in App.~\ref{sec:Coupling_Energies}.
Then, the anisotropic coupling links only states with identical $m$.
To identify excitons based on their main orbital character, we label them according to that $\{l,m\}$ component which produces the largest $\sum_{{\bf k}} \left\vert R_{\lambda,l,m}(k) Y_l^m(\theta,\varphi) \right\vert^2$.
For example, a $1s$ exciton is dominated by a $\{0,0\}$ component.


\begin{figure}[t]
\includegraphics{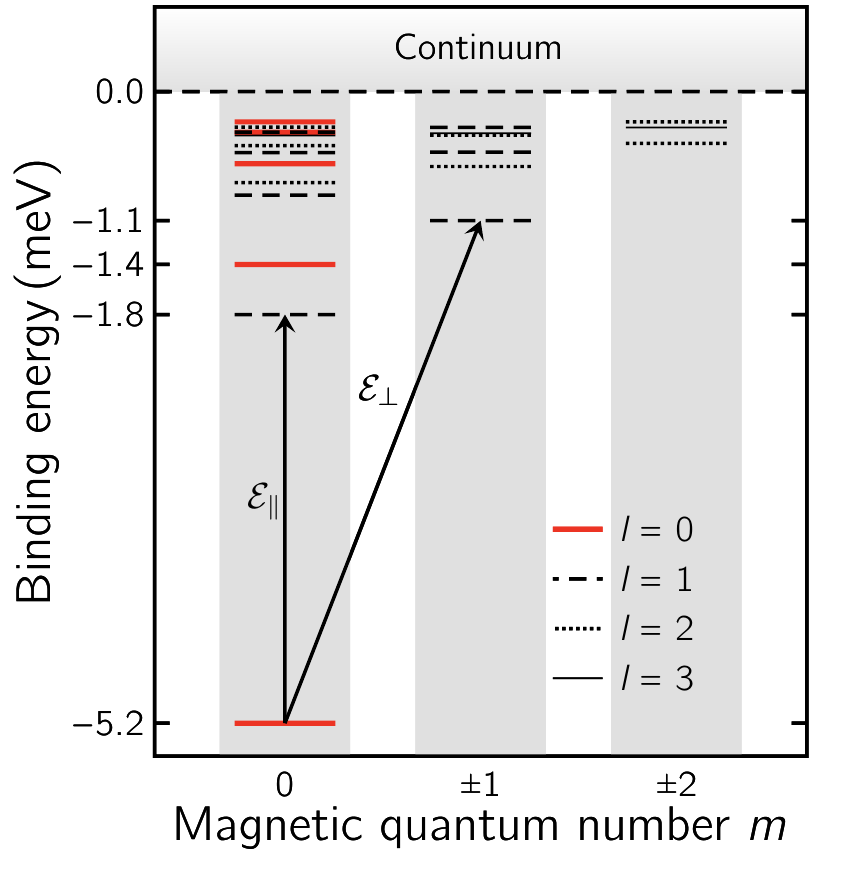}
\caption{
Binding energy of exciton states in Ge corresponding to their dominant character.
Each column represents a magnetic quantum number while the $l$ state are denoted by different line styles.
If excitons are exclusively populated in the ground state, the energetically lowest possible transition for parallel and perpendicular THz field polarizations are marked with arrows.
}
\label{fig:energy_diagram}
\end{figure}

\section{Exciton Mass Anisotropy in Germanium}
\label{sec:analysis}

To demonstrate the overall implications of the mass anisotropy, we consider weak excitations ($f^{\mathrm{e}}_{k} = f^{\mathrm{h}}_{k} = 0$) in Ge.
In practice, we solve Eq.~(\ref{eq:radial_Wannier}) numerically for a finite number of $l$ and $m$ states.
Figure~\ref{fig:numerics}(a) presents the dominant radial parts $R_{\mathrm{GS},l,m}(k)$ of the ground state ($\lambda=0\equiv \mathrm{GS}$), showing that the $\{l=0,m=0\}$ fraction peaks roughly ten times higher than a $\{2,0\}$ component.
Therefore, the ground state exciton has dominantly a $\{0,0\}$ character.
To verify the convergency, Fig.~\ref{fig:numerics}(b) shows the transition energy $E_{\mathrm{GS} \to 1}$ between the two energetically lowest states as a function of the number of included $l$ states.
As we can see, converged results are obtained already for seven $l$ states.

The exciton energy structure is shown in Fig.~\ref{fig:energy_diagram} for five different $m$ states.
For isotropic systems, the first excited $\{l=0,m=0\}$ and the lowest $\{l=1,m=0\}$ and $\{l=1,m=\pm1\}$ states define the energetically degenerate $2s$ and $2p$ states, respectively.
The mass anisotropy clearly removes this degeneracy, producing a $\SI{0.415}{\milli\electronvolt}$ ($\SI{0.362}{\milli\electronvolt}$) separation between the $2s$ and the lowest $\{l=1,m=0\}$ ($\{l=1,m=\pm1\}$) states.

\begin{figure}[tbp]
\includegraphics[]{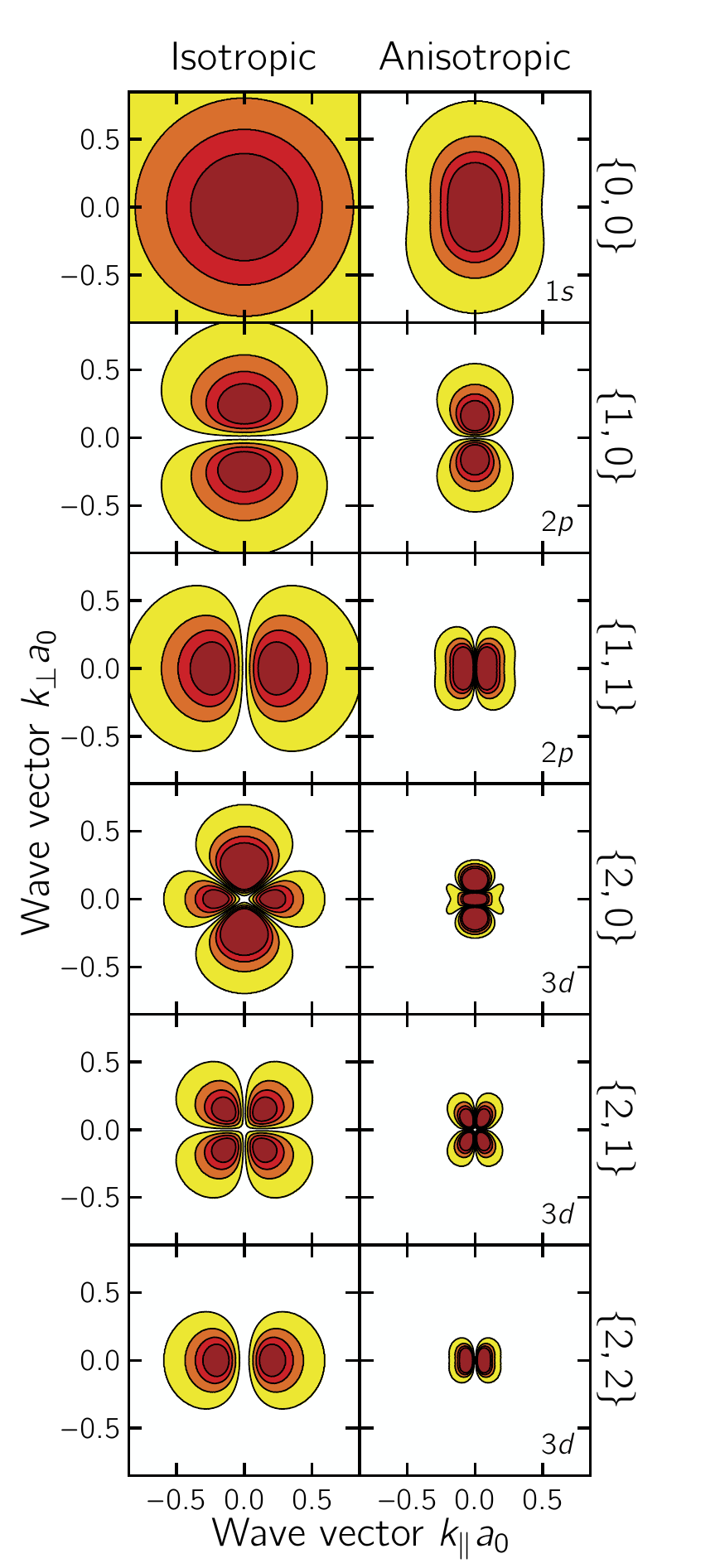}
\caption{
Comparison of the wave functions $\left\vert \phi_\lambda^\mathrm{R}(\bf{k})\right\vert$ for $l \le 2$ using isotropic (left column, $m_\parallel=m_\perp=\SI{1.59}{\Electronmass}$) and anisotropic (right column, $m_\parallel=\SI{1.59}{\Electronmass}$, $m_\perp=\SI{0.0815}{\Electronmass}$) masses.
The dominant character of the individual rows are indicated on the right.
Additionally, the respective symmetry family in more familiar terms $1s$, $2p$, and $3d$ is displayed in each row.
Each wave function is the energetically lowest for its respective dominant symmetry.
}
\label{fig:wave_fcts}
\end{figure}

To visualize the effects of the effective-mass anisotropy, Fig.~\ref{fig:wave_fcts} compares a collection of exciton wave functions $\left\vert \phi^{\mathrm{R}}_\lambda(\bf{k}) \right\vert$ for an isotropic (left column) and an anisotropic (right column) case.
Each wave function represents the energetically lowest state for the dominant symmetry annotated on the right-hand side.
The anisotropic results correspond to Ge masses and the isotropic case is computed with $m_\parallel=m_\perp=\SI{1.59}{\Electronmass}$.
As expected, the isotropic case produces the well-known hydrogen-like wave functions.
Anisotropy already modifies the ground state from a spherically symmetric $s$-wave (isotropic) to a squeezed, slightly peanut-shaped wave function.
The squeezing is explained by the heavier mass in the $k_\parallel$ direction.
The corresponding distortions become significant only in the presence of strong anisotropy.
We also observe that anisotropy squeezes $2s$ (not shown) and $2p$ states to very different shapes compared to their isotropic counterparts, which explains why they become energetically non-degenerate.

\subsection{Single-Valley THz Absorption}
\label{sec:single_valley}

Even though the rapid inter-valley equilibration~\cite{Zhou1994} may make it experimentally difficult to confine the excitons into a single valley, it is instructive to first study the THz response for such a situation and then generalize it for multi-valley excitations, as done in Sec.~\ref{sec:multi_valley}.
Here, we assume that the $L$ valley is located at ${\bf k}_0 = \tfrac{\sqrt{3}\text{{\greektext p}}}{a} {\bf e}_z$ which contains the lattice constant $a$ and is aligned with the $z$-axis as mentioned in Sec.~\ref{sec:theory}.

To study well defined cases, we assume that the THz field is polarized either along the $z$-axis, ${\bf e}_A = {\bf e}_z \equiv {\bf e}_\parallel$, or perpendicular to it, ${\bf e}_A = x {\bf e}_x + y {\bf e}_y \equiv {\bf e}_\perp$ with $x^2 + y^2 = 1$.
The THz response~(\ref{eq:THz_Elliott}) to ${\bf e}_\perp$ polarized light is identical for any $(x,y)$ combination.
Therefore, we may chose $y=0$ and $x=1$ without loss of generality.
The corresponding ${\bf e}_\parallel$ and ${\bf e}_\perp$ polarized fields excite the system via
\begin{align}
j_{\parallel}({\bf k}) = - \frac{e\hbar k_z}{\mu_\parallel} \, ,
\qquad
j_{\perp}({\bf k}) = - \frac{e\hbar k_x}{\mu_\perp} \, ,
\end{align}
which yields the transition-matrix elements
\begin{align}
\label{eq:THz_transition_mel_para}
J_{\lambda,\nu}^\parallel
& =
-\frac{e\hbar}{\mu_z} \int\limits_0^\infty \mathrm{d}k \, k^3 \sum_{l,m} R^\star_{\lambda,l,m}(k) \nonumber \\
& \times
\left[
g_l^m R_{\nu,l+1,m}(k) + g_{l-1}^m R_{\nu,l-1,m}(k)
\right]
\, ,
\end{align}
and
\begin{align}
\label{eq:THz_transition_mel_perp}
J_{\lambda,\nu}^\perp
& =
- \frac{e\hbar}{\mu_x} \int\limits_0^\infty \mathrm{d}k \, k^3 \sum_{l,m} R^\star_{\lambda,l,m} (k) \nonumber \\
& \times
\Big[
h_l^m  R_{\nu,l+1,m+1} (k)
+ h_{l-1}^{m-1} R_{\nu,l-1,m-1} (k)
\nonumber \\
&
- h_{l}^{-m} R_{\nu,l+1,m-1} (k)
- h_{l-1}^{-m-1} R_{\nu,l-1,m+1} (k)
\Big]
\, ,
\end{align}
between the exciton states $\lambda$ and $\nu$ where
\begin{align}
g_l^m & = \sqrt{\frac{(l-m+1) (l+m+1)}{(2l+3) (2l+1)} } \, , \\
h_l^m & = -\frac{1}{2} \sqrt{\frac{(l+m+1)(l+m+2)}{(2l+1)(2l+3)}} \, ,
\end{align}
and the vectorial sum has been replaced by an integral in the usual way~\cite{Haug2009a}.

Equations~(\ref{eq:THz_transition_mel_para}) and~(\ref{eq:THz_transition_mel_perp}) imply different selection rules caused by the mass anisotropy.
In the ${\bf e}_\parallel$ direction, $J_\lambda^\nu$ is non-zero only for transitions involving equal $m$ components while the $l$ components differ by $\pm1$.
For the ${\bf e}_\perp$ polarized exciton, $m$ components couple to $m \pm 1$ and $l$ to $l\pm1$.
Assuming that only the exciton ground state is occupied, the energetically lowest THz transition with non-vanishing transition-matrix element will involve the dominantly $\{0,0\}$ and $\{1,0\}$ states for ${\bf e}_\parallel$ excitation.
However, an ${\bf e}_\perp$ excitation involves $\{0,0\}$ and $\{1,\pm1\}$ components.
Due to the energy difference between the $\{1,0\}$ and $\{1,\pm1\}$ states, the transition energy to the lowest excited state is $\mathcal{E}_\parallel = \SI{3.38}{\milli\electronvolt}$ for ${\bf e}_\parallel$ polarized excitation and $\mathcal{E}_\perp = \SI{4.16}{\milli\electronvolt}$ for ${\bf e}_\perp$ polarized excitation.
Both transitions are illustrated in Fig.~\ref{fig:energy_diagram} by arrows.
Since $\mathcal{E}_\parallel$ and $\mathcal{E}_\perp$ differ by $\SI{0.78}{\milli\electronvolt}$, the mass anisotropy produces direction-dependent THz resonances.
However, a realistic experiment typically contains excitations in all $L$ valleys, which mixes ${\bf e}_\parallel$ and ${\bf e}_\perp$ responses, as discussed in Sec.~\ref{sec:multi_valley}.

\begin{figure}[tbp]
\includegraphics[]{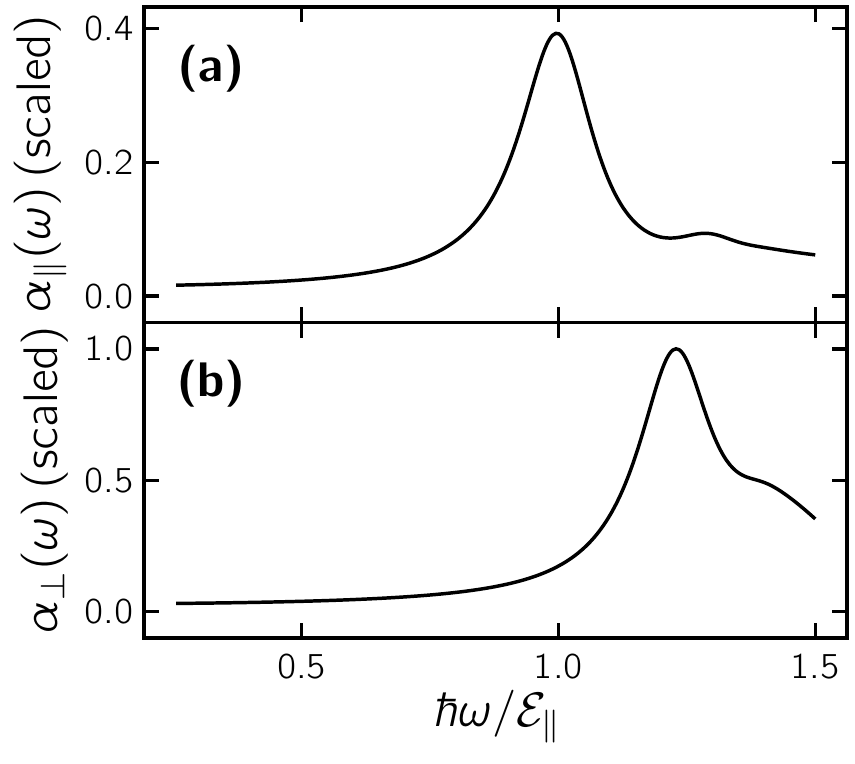}
\caption{
Comparison of the THz absorption spectrum in bulk Ge for parallel (a) and perpendicular (b) polarization at a single $L$ valley.
Excitons are assumed to only populate the ground state.
The energy axis is normalized to the transition energy $\mathcal{E}_\parallel$ of the lowest possible transition for parallel polarization, see Fig.~\ref{fig:energy_diagram}.
}
\label{fig:abs_para_perp}
\end{figure}

A configuration where initially the $1s$ exciton state is populated can be realized at sufficiently low temperatures and long after a weak optical excitation, yielding abundant exciton formation~\cite{Damen1990}.
We then have $n_{\lambda,\nu} = \delta_{\lambda,\nu}\delta_{\lambda,\mathrm{GS}} n_\mathrm{GS}$ where $n_\mathrm{GS}$ is the ground-state exciton density.
Choosing $\gamma = \gamma_J = \SI{0.3}{\milli\electronvolt}$, we compute the THz absorption $\alpha_\parallel(\omega)$ and $\alpha_\perp(\omega)$ for parallel and perpendicular polarizations, respectively, according to Eqs.~(\ref{eq:THz_Elliott})--(\ref{eq:THZ_trans_MEL}), (\ref{eq:THz_transition_mel_para}), and~(\ref{eq:THz_transition_mel_perp}).
Figures~\ref{fig:abs_para_perp}(a) and (b) show $\alpha_\parallel(\omega)$ and $\alpha_\perp(\omega)$, respectively. As we can see, the absorption spectra exhibit two distinctly different resonance energies, corroborating the symmetry-based arguments about the selection rules.
At the same time, the peak heights differ by a factor of $\num{2.54}$, and the parallel direction yields a less asymmetric tail towards high energies.

\begin{figure}[tbp]
\includegraphics[]{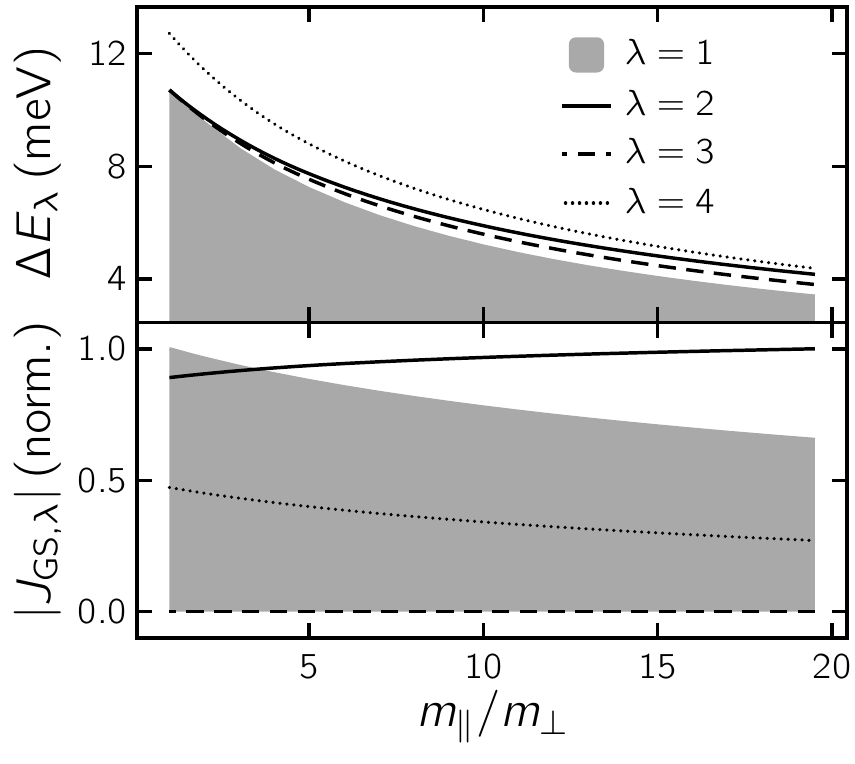}
\caption{
(a) Energetic difference $\Delta E_\lambda = E_\lambda-E_0$ between the four lowest excited states.
(b) THz-transition matrix-element from the ground state to the corresponding end states indicated in (a).
For the end state $\lambda = 3$, neither parallel nor perpendicular polarization allow a transition from the ground state.
Hence, the matrix element vanishes.
}
\label{fig:anisotropy}
\end{figure}

To identify general trends of THz-transition energies and oscillator strengths, we study the binding energies and transition-matrix elements as a function of anisotropy $m_\parallel/m_\perp$ while $m_\parallel$ is kept constant.
Figure~\ref{fig:anisotropy}(a) shows the $\lambda$-to-$\mathrm{GS}$ transition energy $\Delta E_\lambda = E_\lambda-E_\mathrm{GS}$ for the first four excited states.
With vanishing anisotropy ($m_\parallel/m_\perp = 1$), the first three excited excitons ($\lambda = \{1,2,3\}$) become energetically degenerate, as expected for pure $2p$ and $2s$ excitons.
With increasing anisotropy, this degeneracy is lifted.
As a typical trend, $\Delta E_\lambda$ changes fast close to $m_\parallel/m_\perp = 1$ while it saturates with increasing $m_\parallel/m_\perp$.

The saturation can be attributed to the fact that decreasing $m_\perp$ (or increasing $m_\parallel/m_\perp$) freezes the motion in the parallel direction compared to the perpendicular direction.
We also can deduce from the THz-Elliott formula that $J_{\mathrm{GS},\lambda}$ determines the strength of absorption for the $\lambda$-to-$\mathrm{GS}$ transition.
Figure~\ref{fig:anisotropy}(b) shows the transition-matrix elements from the ground to the first four excited states.
The selection rules determine that transitions to $\lambda=1$ and $\lambda=4$ states occur only for the ${\bf e}_\parallel$ polarization.
Conversely, ${\bf e}_\perp$ excitation yields only transitions to the $\lambda=2$ state whereas $\lambda=3$ remains dark for both polarizations because it has the same $s$-like symmetry as the ground state.
As a measure of the overlap of the participating wave functions, $\left\vert J_{\mathrm{GS},\lambda} \right\vert$ monotonically decreases for parallel polarization while it increases in the perpendicular case.
Consequently, the corresponding THz absorption appears stronger in perpendicular polarization as observed in Fig.~\ref{fig:abs_para_perp}.

\begin{figure}[tbp]
\includegraphics[]{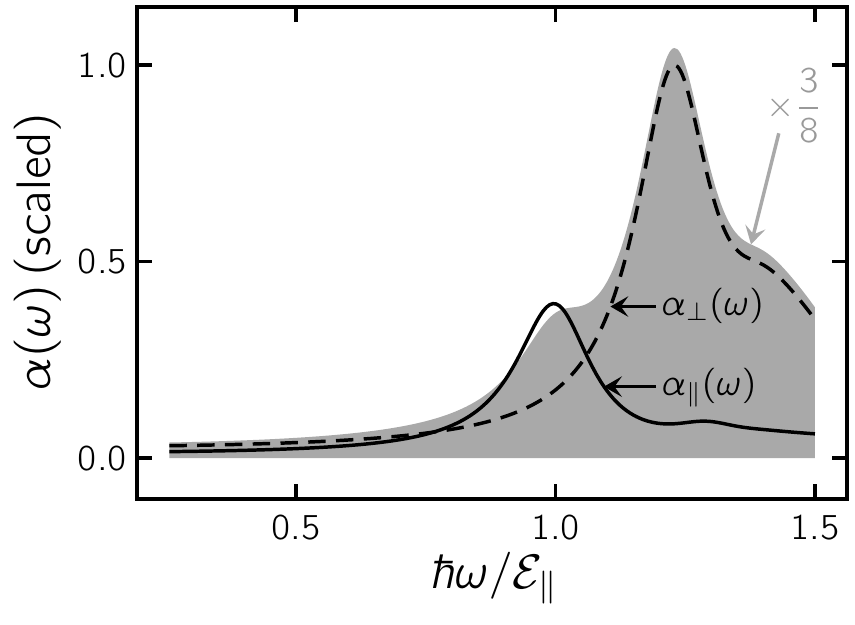}
\caption{
The full absorption spectrum (shaded area) according to Eq.~(\ref{eq:full_abs}) has been scaled by a factor of $3/8$ for visibility.
It is compared to the absorption for parallel (solid) and perpendicular (dashed) polarization from Fig.~\ref{fig:abs_para_perp}.
}
\label{fig:THz_resp}
\end{figure}


\subsection{Multi-Valley THz Absorption}
\label{sec:multi_valley}

Since Ge experiments typically generate excitons in \textit{all} of its $L$ valleys, it is important to consider how this aspect alters the results of Sec.~\ref{sec:single_valley}.
Any arbitrary THz field polarization addresses each $L$ point with a different mixture of perpendicular and parallel directions.
Hence, the total susceptibility that defines the overall absorption must be a tensor.
It is composed out of the individual susceptibility tensors for each valley and diagonal with the components $\chi_{1,1} = \chi_{2,2} = \chi_\perp$ and $\chi_{3,3} = \chi_\parallel$ for the single $L$ valley discussed in Sec.~\ref{sec:single_valley}.
We show in App.~\ref{sec:susc} that not only the total tensor but also its elements remain diagonal, and the overall absorption becomes isotropic
\begin{align}
\label{eq:full_abs}
\alpha(\omega) = \frac{4}{3} \left[
2\alpha_\perp(\omega)
+ \alpha_\parallel(\omega)
\right] \, .
\end{align}
As a simple interpretation of Eq.~(\ref{eq:full_abs}), $1/3\left[ 2\alpha_\perp(\omega) + \alpha_\parallel(\omega)\right]$ defines the mean over three cartesian directions, and the factor four accounts for the absorption of the four unique $L$ valleys.

Figure~\ref{fig:THz_resp} compares the total THz absorption (shaded area) with its $\alpha_\parallel$ (solid line) and $\alpha_\perp$ (dashed line) constituents.
Even though $\alpha(\omega)$ is a mixture of all directions, it still exhibits a residual double-peak structure due to the mass anisotropy.
Therefore, a THz measurement in Ge can directly detect the mass anisotropy in THz absorption spectra as long as the dephasing does not greatly exceed the anisotropic splitting ($\SI{0.78}{\milli\electronvolt}$).


\section{Conclusions}
\label{sec:conclusions}

We have presented an efficient numerical method to solve the Wannier equation for systems with mass anisotropy based on an expansion into spherical harmonics.
We use Ge as a prototype system to illustrate the main consequences of the mass anisotropy for the exciton wave functions and the THz absorption.
We find that the anisotropy lifts the degeneracy of excitons with the same $l$ quantum number.
We find excitonic wave function modifications that can be traced back to the coupling of different $l$ and $m$ components in our expansion in terms of spherical harmonics.

The wave function distortions result in modified selection rules for THz induced transitions, depending on the polarization of the applied THz field.
For polarization parallel to the $\varGamma \to L$ direction, transitions involve states whose $l$ differs by $\pm1$ while $m$ remains constant.
For perpendicular polarizations, we find that both $l$ and $m$ are changed by $\pm 1$ between the states.
As a consequence, two distinct THz resonances can be observed, each assigned to one of the polarization directions.
Our calculations indicate that these resonances should be observable in good quality
THz-absorption experiments.

Besides excitonic features, future experiments could also explore how mass anisotropy modifies properties of more complicated quasiparticles such as biexcitons~\cite{Vektaris1994,Chen2002} and dropletons~\cite{Almand-Hunter2014}.
In particular, the presented approach can be used as a starting point to also compute the relevant correlation functionals to determine the energetically possible configurations~\cite{Mootz2013}.


\begin{acknowledgments}
This work is funded by the German Science Foundation (DFG) through the Research Training Group \textit{Functionalization of Semiconductors} (GRK 1782) and via Grants No. KI 917/3-1 and No. SFB 1083.
\end{acknowledgments}


\appendix


\section{Generalized Wannier Equation for Indirect Excitons}
\label{sec:Wannier_Equation_for_Indirect_Excitons}

Analogous to direct semiconductors, the exciton wave function in an indirect semiconductor is determined by the homogenous solution to the semiconductor Bloch equations (SBE)~\cite{Lindberg1988}.
However, we now have to allow for transitions between different momentum states of the participating carriers.
Repeating the derivation of the SBE in Hartree--Fock approximation for the interband polarization $
p_{{\bf k},{\bf k}'} = \left<
\hat{a}_{\mathrm{v},{\bf k}\phantom{'}}^{\dagger}
\hat{a}_{\mathrm{c},{\bf k}'}
\right>
$, the homogeneous solution reads
\begin{align}
\label{eq:Wannier_ind_appendix}
E_\lambda \phi_\lambda^{\mathrm{R}}({\bf k}, {\bf k}')
& =
\tilde{E}_{{\bf k},{\bf k}'} \phi_\lambda^{\mathrm{R}}({\bf k}, {\bf k}')
\nonumber \\
& - \left(
1 - f^{\mathrm{e}}_{{\bf k}'} - f^{\mathrm{h}}_{{\bf k}}
\right)
\sum\limits_{{\bf q}}
V_{q}
\phi_\lambda^{\mathrm{R}}({\bf k}-{\bf q}, {\bf k}'-{\bf q}) \, ,
\end{align}
in which the electron--hole pair energy is defined as
\begin{equation}
\label{eq:ren_e_h_ene_indir_appendix}
\tilde{E}_{{\bf k},{\bf k}'}
=
E^{\mathrm{e}}_{{\bf k}'} + E^{\mathrm{h}}_{{\bf k}} - E_\mathrm{g}
- \sum\limits_{{\bf q}}
V_{q}
\left(
f^{\mathrm{e}}_{{\bf k}' - {\bf q}} + f^{\mathrm{h}}_{{\bf k} - {\bf q}}
\right)
\, .
\end{equation}
The solutions of Eq.~(\ref{eq:Wannier_ind_appendix}) define the wave functions of an indirectly bound electron--hole pair and their binding energies.
To further simplify the problem, we introduce the concept of \textit{diagonal excitons}, e.g.~the momentum difference is assumed to be constant: ${\bf k}' = {\bf k} + {\bf k}_0$.
Equations~(\ref{eq:Wannier_ind_appendix}) and~(\ref{eq:ren_e_h_ene_indir_appendix}) then reduce to Eqs.~(\ref{eq:Wannier}) and~(\ref{eq:ene_disp_aniso}), respectively.


\section{Coupling Strengths}
\label{sec:Coupling_Energies}

The coupling strengths in Eq.~(\ref{eq:radial_Wannier}) emerge when the kinetic part of the Wannier equation is projected on to spherical harmonics, e.g.~by computing
\begin{align}
\mathfrak{I}_{\bf k}
\equiv
\int \mathrm{d}\varOmega
\left[
Y_{l}^{m}(\varOmega)
\right]^\star
E_{\bf k} \phi_\lambda^\mathrm{R}(\bf k) \, ,
\end{align}
where the unrenormalized part of Eq.~(\ref{eq:ene_disp_aniso}) in spherical coordinates is
\begin{align}
E_{\bf k} &= \frac{\hbar^2k^2}{2}
\nonumber \\
&\times
\left(
\frac{\sin^2 (\theta) \cos^2 (\varphi)}{\mu_x}
+ \frac{\sin^2 (\theta)\sin^2 (\varphi)}{\mu_y}
+ \frac{\cos^2 (\theta)}{\mu_z}
\right) \, ,
\end{align}
and the solid angle is given by $\varOmega \equiv (\theta,\varphi)$.
The recurrence formula for the associated Legendre polynomials~\cite{Abramowitz1972} are very useful to solve these integrals.
Defining the masses
\begin{align}
\frac{1}{\mu^{\pm}_{xy}} = \frac{1}{\mu_x} \pm \frac{1}{\mu_y} \, , \qquad
\frac{1}{\mu^{\pm}_{xyz}} = \frac{1}{\mu^{+}_{xy}} \pm \frac{2}{\mu_z} \, ,
\end{align}
leads to the following coupling strengths
\begin{align}
\epsilon_{l,m}^{(1)} & = N_l \left( \frac{\mu_z}{\mu^{+}_{xyz}} L_l + \frac{\mu_z}{\mu^{-}_{xyz}} M_m - 3 \right) \, , \nonumber \\
\epsilon_{l,m}^{(2)} & = -\frac{\mu_z}{\mu^{-}_{xyz}} \frac{\sqrt{N_l} J_{l,m}}{2(2l+1)} \, , \nonumber \\
\epsilon_{l,m}^{(3)} & = -\frac{\mu_z}{\mu^{-}_{xy}} \frac{N_l J_{l,m}}{2} \, , \nonumber \\
\epsilon_{l,m}^{(4)} & = \frac{\mu_z}{\mu^{-}_{xy}} \frac{\sqrt{N_l}}{4(2l+1)} B_{l,m} \, ,
\end{align}
containing the coefficients
\begin{align}
N_l & = \frac{1}{(2l-1)(2l+3)} \, , \quad
L_l = l(l+1) \, , \nonumber \\
M_m & = (m+1)(m-1) \, , \quad
B^2_{l,m} = {\prod\limits_{j=0}^3 (l+m+j)} \, , \nonumber \\
J_{l,m} & = \sqrt{
\Big[
(l+1)^2-m^2
\Big]
\Big[
l^2-m^2
\Big]
} \, .
\end{align}


\section{Coulomb-Matrix Element}
\label{sec:Coulomb_Matrix_Element}

Inserting Eq.~(\ref{eq:WF_ansatz}) into the Coulomb part of Eq.~(\ref{eq:Wannier}) yields
\begin{align}
\label{eq:I_k_1}
I_{{\bf k}}
&\equiv
\sum_{{\bf k}'} V_{|{\bf k} - {\bf k}'|} \phi_\lambda^\mathrm{R}({\bf k}') \nonumber \\
&=
\left(
\tfrac{L}{2\text{{\greektext p}}}
\right)^3
\int \mathrm{d}^3 k' \, V_{|{\bf k} - {\bf k}'|} Y_l^m(\varOmega') \nonumber \\
&= \sum_{l,m} \int\limits_0^\infty \mathrm{d}k' \left(k'\right)^2 R_{n,l,m}(k')
\nonumber \\
&\times
\int \mathrm{d}\varOmega'
\left(
\tfrac{L}{2\text{{\greektext p}}}
\right)^3
V_{|{\bf k} - {\bf k}'|} Y_l^m(\varOmega')
\, ,
\end{align}
where we have first replaced the vectorial sum by an integral which is then implemented in spherical coordinates ${\bf k} = (k,\varOmega)$.
To evaluate the solid angle integral, we need the explicit form of the Coulomb potential
\begin{align}
\label{eq:Coulomb_bulk}
V_{|{\bf k} - {\bf k}'|} =
\frac{1}{L^3} \int \mathrm{d}^3 r \,
V(r) \mathrm{e}^{-\mathrm{i} {\bf k} \cdot {\bf r}} \mathrm{e}^{\mathrm{i} {\bf k}' \cdot {\bf r}} \, ,
\end{align}
where $V(r) = \tfrac{e^2}{4\text{{\greektext p}}\epsilon_0\epsilon_\mathrm{r} r}$ contains the background dielectric constant $\epsilon_\mathrm{r}$, and $\epsilon_\mathrm{r} = 16$ in Ge~\cite{Madelung1991}.
Using Eq.~(\ref{eq:Coulomb_bulk}) and the plane-wave expansion~\cite{Mehrem2011}
\begin{align}
\mathrm{e}^{-\mathrm{i} {\bf k} \cdot {\bf r}} =
4\text{{\greektext p}} \sum_{l,m}
(-\mathrm{i})^l j_l^\star(kr)
\left[Y_l^m(\varOmega)
\right]^\star
Y_l^m(\varTheta) \, ,
\end{align}
with ${\bf r} = (r,\varTheta)$, the $\Omega'$ integral in Eq.~(\ref{eq:I_k_1}) can be expressed via
\begin{align}
\label{eq:Define_Coulomb}
\int\mathrm{d}\varOmega' \left(\tfrac{L}{2\text{{\greektext p}}}\right)^3 V_{|{\bf k} - {\bf k}'|} Y_l^m(\varOmega') \equiv Y_l^m(\Omega) \mathcal{V}^{l}_{k,k'} \, ,
\end{align}
where
\begin{align}
\label{eq:I_k_2}
\mathcal{V}^{l}_{k,k'} = \frac{2}{\text{{\greektext p}}}\int\limits_0^\infty \mathrm{d}r \, r^2V(r) j_l^\star(kr)j_l(k'r) \, ,
\end{align}
containing the spherical Bessel functions $j_l$.
In this format, the projection of Eq.~(\ref{eq:I_k_1}) to spherical harmonics is particularly easy since
\begin{align}
\int \mathrm{d}\varOmega \left[Y_{l'}^{m'}(\varOmega)\right]^\star I_{\bf k}
& = \int\limits_0^\infty \mathrm{d}k' \left(k'\right)^2 R_{n,l,m}(k') \mathcal{V}^{l}_{k,k'} \, ,
\end{align}
due to the orthonormality of $Y_l^m$.
At the same time, Eq.~(\ref{eq:Define_Coulomb}) yields
\begin{align}
\mathcal{V}^{l}_{k,k'} = \int \mathrm{d}\varOmega' \left(\tfrac{L}{2\text{{\greektext p}}}\right)^3 V_{|{\bf k} - {\bf k}'|} \frac{Y_l^m(\varOmega')}{Y_l^m(\varOmega)} \, ,
\end{align}
where the right-hand side must also be independent of $m$ and $\varOmega$.
Hence, we can chose them at our convenience.
The simplest choice $m=0$ and $\theta=\varphi=0$ yields
\begin{equation}
\label{eq:Coul_angle_ave}
\mathcal{V}^{l}_{k,k'} = \frac{e^2}{4\text{{\greektext p}}^2\epsilon_0\epsilon_\mathrm{r}} \int\limits_0^\text{{\greektext p}} \mathrm{d}\theta' \frac{\sin(\theta') P_l(\cos(\theta'))}{k^2+\left(k'\right)^2 - 2kk'\cos(\theta')} \, ,
\end{equation}
with the Legendre polynomials $P_l$.
Not having to introduce an additional numerical grid is the main advantage of Eq.~(\ref{eq:Coul_angle_ave}) over Eq.~(\ref{eq:I_k_2}).
We note that $P_l$ can always be written in the form $P_l(x) = \sum_{j=0}^l a_{l,j} x^j$ with the expansion coefficients~\cite{Abramowitz1972}
\begin{align}
a_{l,j} = 2^l \binom{l}{j}\binom{\tfrac{1}{2}(l+j-1)}{l} \, .
\end{align}
Thus, the remaining integral in Eq.~(\ref{eq:Coul_angle_ave}) becomes
\begin{align}
\label{eq:Coulomb_int}
\mathbb{I}_j & \equiv
\frac{1}{K} \int\limits_0^\text{{\greektext p}} \mathrm{d}\cos(\theta') \frac{\cos^j(\theta')}{1-Q\cos(\theta')}
\nonumber \\
&=
\frac{2}{K Q^{j+1}} \sum\limits_{n=p+1}^\infty \frac{Q^{2n-1}}{2n-1}
 \, ,
\end{align}
with $K \equiv k^2+\left(k'\right)^2$, $0 \leq Q \equiv \tfrac{2kk'}{K} \leq 1$, and $p=j/2$ ($p=(j+1)/2$) if $j$ is even (odd).
For high orders in $j$ and small $Q$, $\mathbb{I}_j$ can show numerical instabilities due to limited machine precision.
In this case, the series in Eq.~(\ref{eq:Coulomb_int}) has to be terminated approproiatly.
Finally, the Coulomb renormalization of the kinetic part of Eq.~(\ref{eq:Wannier}),
\begin{align}
\mathcal{I}_{\bf k}
\equiv
\sum\limits_{{\bf k}'}
V_{|{\bf k} - {\bf k}'|}
\left(
f^{\mathrm{e}}_{{\bf k}'} + f^{\mathrm{h}}_{{\bf k}'}
\right) \, ,
\end{align}
is projected to spherical harmonics again by substituting the vectorial sum with an integral in spherical coordinates.
If we make use of the freedom to define ${\bf k} = k{\bf e}_z$ under the integral, we obtain
\begin{align}
\int \mathrm{d}\varOmega \left[Y_{l'}^{m'}(\varOmega)\right]^\star \mathcal{I}_{\bf k}
& = \int\limits_0^\infty \mathrm{d}k' \left(k'\right)^2
\nonumber \\
&\times
\left(
f^{\mathrm{e}}_{k'} + f^{\mathrm{h}}_{k'}
\right) 
\mathcal{V}^0_{k,k'} R_{\lambda,l,m}(k) \, ,
\end{align}
if we assume radially symmetric carrier distributions.


\section{Susceptibility Tensor}
\label{sec:susc}

The macroscopic current can be decomposed as
\begin{align}
{\bf J}
=
\sum_{n} {\bf J}_{n}
=
\sum_{n} \sum_{\lambda,{\bf k}_n} {\bf j}^\lambda_{{\bf k}_n} \hat{a}^{\dagger}_{\lambda,{\bf k}_n} \hat{a}^{\phantom{\dagger}}_{\lambda,{\bf k}_n} \, ,
\end{align}
where ${\bf k}_n$ is in the vicinity of the $n$-th $L$ valley.
The center of the four unique $L$ valleys are
\begin{align}
{\bf L}_1
&=
\begin{pmatrix}
0 & 0 & 1
\end{pmatrix}
\, ,
&{\bf L}_2
&=
\frac{1}{3}
\begin{pmatrix}
-\sqrt{2} & -\sqrt{6} & 1
\end{pmatrix}
\, , \nonumber \\
\label{eq:directions}
{\bf L}_3
&=
\frac{1}{3}
\begin{pmatrix}
-2\sqrt{2} & 0 & -1
\end{pmatrix}
\, ,
&{\bf L}_4
&=
\frac{1}{3}
\begin{pmatrix}
-\sqrt{2} & \sqrt{6} & 1
\end{pmatrix}
\, .
\end{align}
Let ${\bf e}_{n,j}$ ($j=\{x,y,z\}$) denote the cartesian unit vectors where ${\bf e}_{n,z}$ is aligned with ${\bf L}_n$.
Then, the tensorial susceptibility becomes $\chi_\perp({\bf e}_{1,x} \otimes {\bf e}_{1,x} + {\bf e}_{1,y} \otimes {\bf e}_{1,y}) + \chi_\parallel{\bf e}_{1,z} \otimes {\bf e}_{1,z}$ for the valley with $n=1$.
For all valleys, we find
\begin{align}
\overline{\overline{\chi}}(\omega)
&=
\sum_n
\big[
\chi_\perp
\left({\bf e}_{n,x} \otimes {\bf e}_{n,x} + {\bf e}_{n,y} \otimes {\bf e}_{n,y}
\right)
\nonumber \\
&+ \chi_\parallel{\bf e}_{n,z} \otimes {\bf e}_{n,z}
\big]
=
\frac{4}{3} (2\chi_\perp + \chi_\parallel) \mathbb{1} \, ,
\end{align}
where the last step follows as the explicit directions~(\ref{eq:directions}) are applied.


\bibliography{library}

\end{document}